\documentclass[iop]{emulateapj}
\usepackage{graphicx,epsfig,fancyhdr,rotating,amsmath} 
\usepackage{txfonts}
\usepackage{epsfig,rotate,morefloats}
\usepackage[breaklinks,colorlinks,urlcolor=blue,citecolor=blue,linkcolor=blue]{hyperref}

\shorttitle{}
\shortauthors{Gupta}
\begin{document}
\title{Spectroscopic evidence of Alfv\'en wave damping  in the off-limb solar corona}
\author{G.~R. Gupta}
\affil{Inter-University Centre for Astronomy and Astrophysics, Post Bag-4, Ganeshkhind, Pune 411007, India}
\email{e-mail: girjesh@iucaa.in}

\begin{abstract}

We investigate off-limb active region and quiet Sun corona using spectroscopic data. 
Active region is clearly visible in several spectral lines formed in the temperature range of 1.1--2.8 MK.
We derive electron number density using line ratio method,
and non-thermal velocity in the off-limb region up to the distance of 140 Mm. We compare density
scale heights derived from several spectral line pairs with expected scale heights as per hydrostatic
equilibrium model. Using several isolated and unblended spectral line profiles, we
estimate non-thermal velocities in active region and quiet Sun. Non-thermal velocities obtained 
from warm lines in active region first show increase and later show either decrease or almost 
 constant value with  height in the far off-limb region, whereas hot lines show consistent decrease.
However, in the quiet Sun region, non-thermal velocities obtained from various spectral lines show
either gradual decrease or remain almost constant with height. Using these obtained
parameters, we further calculate Alfv\'en wave energy flux in the both active and
quiet Sun regions. We find significant decrease in wave energy fluxes with height,
and hence provide evidence of Alfv\'en wave damping.  Furthermore, we derive damping lengths of 
Alfv\'en waves in the both regions and find them to be in the range of 25-170 Mm. 
Different damping lengths obtained at different temperatures may be explained as 
either possible temperature dependent damping or measurements obtained in different coronal
structures formed at different temperatures along the  line-of-sight. 
Temperature dependent damping may suggest some role of thermal conduction 
in the damping of Alfv\'en waves in the lower corona.

\end{abstract}


\section{Introduction}\label{intro}

Heating of solar atmosphere and acceleration of solar wind remain two of  the most puzzling problems in the solar and space physics.
There are several theories proposed to explain the phenomena, however, to identify any one dominant process is extremely difficult to do.
For details, see \citet{2012RSPTA.370.3217P}, and \citet{2015RSPTA.37340269D} and references therein for current progress in the field.
Most of the  models proposed so far are attributed to either dissipation of magnetohydrodynamics (MHD) waves or magnetic reconnection.
Among the several proposed ideas, role of wave turbulence in the heating of solar corona and acceleration of solar wind is one of the well
studied model  \citep[see recent reviews by][]{2015RSPTA.37340261A,2015RSPTA.37340148C}. 
\citet{1942Natur.150..405A} first suggested existence of electromagnetic-hydrodynamic waves in the solar atmosphere and its importance
in the heating of solar corona \citep{1947MNRAS.107..211A}. This led to the wave heating model of solar corona. 
In this model, convective motions at the footpoints of magnetic flux tubes are assumed to generate wave-like fluctuations that propagate up into the
extended corona \citep{2005ApJS..156..265C,2005ApJ...632L..49S}.  These fluctuations  are often proposed to partially reflect back down toward the Sun, 
develop into strong magnetohydrodynamics (MHD) turbulence, and dissipate gradually \citep{2007ApJS..171..520C,2010ApJ...708L.116V}.
Recently, \citet{2011ApJ...736....3V} developed a three-dimensional magnetohydrodynamics (MHD) Alfv\'en wave turbulence model to explain the heating of
both solar chromosphere and corona in the coronal loop. Another model used to explain coronal heating is nanoflare heating model \citep[see recent review
by][]{2015RSPTA.37340256K}. In this model, random photospheric motions and flows lead to twisting and braiding of coronal field lines. This
results in  building up of magnetic stress, and thus, leads to release of energy in the form of impulsive heating events called as nanoflares
\citep{1988ApJ...330..474P}.

In order to understand wave heating mechanism of solar atmosphere, observations of detection, propagation, and dissipation of waves are essential.
\citet{2007Sci...317.1192T} and \citet{2011Natur.475..477M}  reported the ubiquitous presence of outward propagating Alfv\'enic (transverse) waves in the solar corona.
Propagating Alfv\'enic waves were also found in the polar region \citep{2010ApJ...718...11G,2015NatCo...6E7813M}.
Comprehensive review exists on the detection of propagating waves in the solar atmosphere, e.g., \citet{2011SSRv..158..267B,2012RSPTA.370.3193D,2015SSRv..190..103J}.
In recent studies, evidence of damping of propagating waves are also reported \citep{2014ApJ...784...29M,2014A&A...568A..96G,2014ApJ...789..118K}.
Signatures of Alfv\'en waves can also be found through the study of broadening of spectral line profiles in the solar corona \citep[e.g.,][]{2009A&A...501L..15B,2009Sci...323.1582J}.
Alfv\'enic wave motions are transverse to the direction of propagation. In case of field lines aligned in the plane of sky, 
 plasma motions due to Alfv\'enic waves will either be directed towards or away from the line-of-sight. In the off-limb corona,
 several spatially unresolved structures may be present along the line-of-sight with different phases of oscillations. These unresolved wave motions can 
lead to non-thermal broadening of spectral line profiles. Thus, observed non-thermal broadening of spectral line profiles in the corona will be
proportional to Alfv\'en wave amplitude, e.g., \citet{2001A&A...374L...9M}.

There are numerous studies devoted to measure off-limb non-thermal broadening of spectral lines to search for any wave activity. 
\citet{1990ApJ...348L..77H} performed the first observations of high temperature line profiles in the solar off-limb region using sounding rocket experiments.
They found increase  in line width with height above the limb and interpreted as a signature of propagating hydromagnetic waves in the solar corona.
Later, more studies were carried out using space based SUMER instrument on-board SOHO. Using SUMER, \citet{1998SoPh..181...91D} and \citet{1998A&A...339..208B}
 found increase in non-thermal line width with off-limb height and associated density decrease. Their results were in excellent agreement with predictions from 
 outward  propagating undamped  Alfv\'en waves. \citet{2002A&A...392..319H} performed similar analysis on off-limb part of quiet Sun corona using CDS instrument
 on-board SOHO. They found narrowing of line width with height and interpreted as indication of wave dissipation in a closed loop system in the low corona.
\citet{2009A&A...501L..15B} performed similar analysis on polar plume and interplume region using Extreme-Ultraviolet Imaging Spectrometer
\citep[EIS, ][]{2007SoPh..243...19C} on-board Hinode \citep{2007SoPh..243....3K}. They found signatures of outward propagating undamped linear Alfv\'en waves
within $1.1~R_{\odot}$. Recently, \citet{2012ApJ...751..110B} and \citet{2012ApJ...753...36H} measured non-thermal line width 
up to $1.4~R_{\odot}$ in the open magnetic field of polar regions using EIS/Hinode. They found signature of damping of Alfv\'en waves beyond  $1.1-1.14~R_{\odot}$.
\citet{2014ApJ...780..177L} investigated cool loop and dark lane over a off-limb active region and obtained basic plasma parameters as a function of height above 
the limb. They found slight decrease in non-thermal velocity along the cool loop whereas sharp fall along the dark lane. They attributed these findings to wave damping.
\citet{2014ApJ...795..111H} also measured energy and dissipation of Alfv\'enic waves in the quiet Sun region.

Recently, \citet{2011ApJ...736....3V}  developed a 3-D MHD model of Alfv\'en wave turbulence  to explain the heating of  solar chromosphere and corona in 
the coronal loop. This model has attracted lot of attention from the community to look for such signatures \citep[e.g.][]{2014ApJ...786...28A}.
 In this work, we focus on off-limb active region loop system and quiet Sun corona to study propagation of Alfv\'en waves  with height and search
 for any signature of their damping  over a wide range of temperature. As Alfv\'en wave energy flux density is given by \citep[e.g.][]{2001A&A...374L...9M},

\begin{equation}
 E_D=\rho \xi^2 V_A=\sqrt{\frac{\rho}{4\pi}}~\xi^2~B
 \label{eq:energyd}
\end{equation}

where $\rho$ is mass density ($\rho=m_p N_e$, $m_p$ is proton mass, and $N_e$ is electron number density), $\xi$ is Alfv\'en wave velocity amplitude,
and $V_A$ is Alfv\'en wave propagation velocity given as $B/\sqrt{4\pi \rho}$. Therefore, total wave energy flux crossing a surface area A will be given by,

\begin{equation}
 E_F= \frac{1}{\sqrt{4\pi}}~\sqrt{m_p N_e}~\xi^2~B~A 
\label{eq:energyf}
\end{equation}

\begin{figure}[htbp]
\centering
\includegraphics[width=8cm]{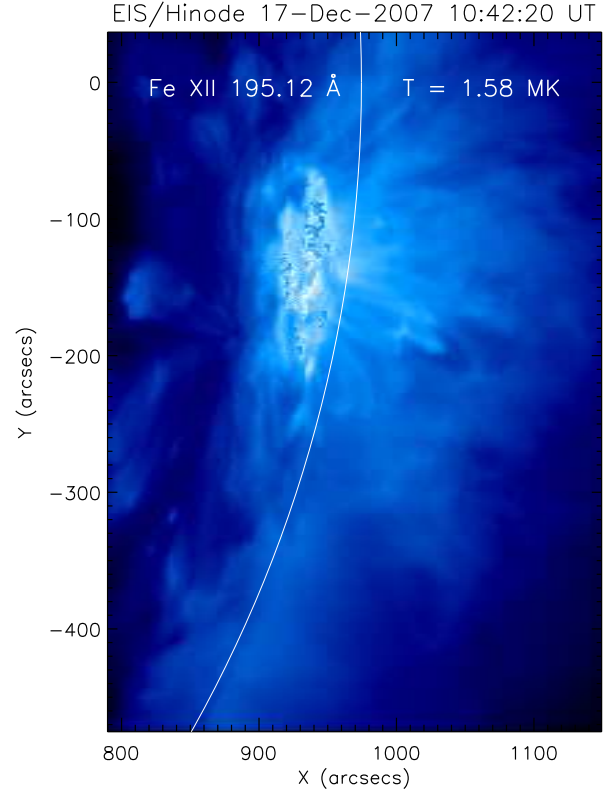}
\caption{Intensity map of off-limb active region and quiet Sun in Fe {\sc xii} 195.12 \AA\  spectral line rastered by EIS/Hinode on 17 December 2007.
Continuous white  line indicates location of solar limb.}
\label{fig:context1}
\end{figure}

Henceforth, total Alfv\'en wave energy flux depends on electron number density, wave amplitude, magnetic field, and area of cross section.
In this paper, our main focus is to estimate total wave energy flux with height in the off-limb solar corona, and thus, to find any signatures of wave damping.
For the purpose, we identified a unique set of good spectroscopic data covering the off-limb active region and quiet Sun observed by EIS/Hinode. 
Data covers various spectral lines formed over a wide range of temperature. Previous such studies  were mainly carried out with few spectral lines
formed at very similar temperature e.g. Fe {\sc xii}, and Fe {\sc xiii}. Therefore, current study provides an unique opportunity to carry out 
such analysis for coronal structures formed over a wide range of temperature. This may also enable us to find any possible existence of temperature dependence.
Related details of observations are described in \S~\ref{sec:obs}. We employ spectroscopic methods to obtain  electron number density,
and non-thermal velocity which are described in \S~\ref{sec:intdens}, and \S~\ref{sec:ntv} respectively. In \S~\ref{sec:energy}, 
we describe calculation of Alfv\'en wave energy flux using obtained parameters. Obtained results are discussed in \S~\ref{sec:discussions}, 
and final summary and conclusions are provided in \S~\ref{sec:conclusion}.

\section{Observations and Data Analysis}\label{sec:obs}

Off-limb active region AR 10978 was observed by EIS/Hinode on 17 December 2007. EIS observations were carried out with 2\arcsec\ slit
and exposure time of 45~s. Observations were performed over the wavelength range of 180--204 \AA\ and 248--284 \AA .
Raster scan started at 10:42:20~UT and completed at 13:02:17~UT and covered a field of view of $360\arcsec \times 512\arcsec$. 
This dataset was previously analyzed by \citet{2011A&A...525A.137O} to study electron density and temperature structure of a limb active region.
We followed standard procedures for preparing the EIS data  using IDL routine 
EIS\_PREP\footnote{\url ftp://sohoftp.nascom.nasa.gov/solarsoft/hinode/eis/doc/eis\_notes/01\_EIS\_PREP/eis\_swnote\_01.pdf}
available in the \textsl{Solar Software} \citep[SSW;][]{1998SoPh..182..497F}.  
Recently, \citet{2016ApJ...820...63B} and \citet{2016ApJ...827...99T} showed that absolute calibration of EIS data leads to a systematic overestimation
of spectral line widths for most of the pixels along slit. Thus, for the purpose of measuring line widths, we obtained EIS spectra in the data number (DN) unit by
applying EIS\_PREP routine with /noabs keyword. Moreover, we also obtained EIS spectra in the physical units (erg cm$^{-2}$ s$^{-1}$ sr$^{-1}$)
 to further perform electron number density diagnostics. Routine also provides errorbars on the obtained intensities.
 In addition, there also exists 22\% uncertainty in the observed intensity based on pre-flight
 calibration of EIS \citep{2006ApOpt..45.8689L}.  All the EIS spectral line profiles were fitted with Gaussian function using 
 EIS\_AUTO\_FIT\footnote{\url ftp://sohoftp.nascom.nasa.gov/solarsoft/hinode/eis/doc/eis\_notes/16\_AUTO\_FIT/eis\_swnote\_16.pdf}.
 Routine also provides one-sigma errorbars on the fitted parameters. Comparison between both type of spectra 
 reconfirms the systematic overestimation of line widths from absolutely calibrated data as recently reported by \citet{2016ApJ...820...63B}
 and \citet{2016ApJ...827...99T}. However, magnitude of this systematic overestimation of line-widths were found to be very small in the current dataset.
  As EIS sensitivity is evolving over time, absolutely calibrated data 
 (in physical units) and related errors were further recalibrated using the method of  \citet{2014ApJS..213...11W}.
  There exists spatial offsets in the solar-X and solar-Y directions between images obtained from different wavelengths.
These offsets were corrected with respect to image obtained from Fe~{\sc xii} 195.12 \AA\ spectral 
 line\footnote{\url ftp://sohoftp.nascom.nasa.gov/solarsoft/hinode/eis/doc/eis\_notes/03\_GRATING\_DETECTOR\_TILT/eis\_swnote\_03.pdf}.  
Figure~\ref{fig:context1} shows intensity map of observed off-limb active region obtained from Fe {\sc xii} 195.12 \AA\ line. 
Observed active region is very bright and has several saturated image pixels at few locations.

\begin{figure}[htbp]
\centering
\includegraphics[width=8.5cm]{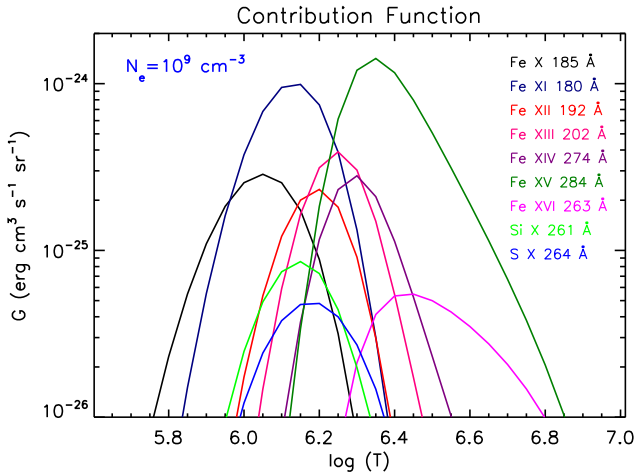}
\caption{Contribution function of spectral lines selected for detailed analysis of off-limb active and quiet Sun regions recorded by EIS/Hinode (see also Table~\ref{tab:lines}).}
\label{fig:cont_func}
\end{figure}

To identify spectral line wavelengths and corresponding peak formation temperatures, all the atomic data used in this study are taken
from CHIANTI atomic database \citep{1997A&AS..125..149D,2015A&A...582A..56D}. To perform line width analysis, we identified several 
unblended and isolated spectral lines with good signal strength as highlighted by \citet{2007PASJ...59S.857Y} (see Table~\ref{tab:lines}).
Although there exist some blend in Fe~{\sc xiv} 274 \AA , and Fe~{\sc xv} 284 \AA\ lines, their contribution can safely be ignored in the active region conditions.
Lines are chosen in such a way to get good coverage over temperature range. Contribution function of selected spectral lines  were 
calculated using the CHIANTI v.8 \citep{2015A&A...582A..56D}  at constant electron number density $N_e=10^9$~cm$^{-3}$.
Obtained contribution function curves are  plotted in Figure~\ref{fig:cont_func}. Peak formation temperature of all the selected
lines are also provided in Table~\ref{tab:lines}. We identify all the spectral lines formed below 2 MK temperature as warm lines
whereas those formed above 2 MK as hot lines.  We also identified several density sensitive lines and utilized them only for the 
purpose of deriving electron number density. 

\begin{table}[thbp]
\centering
\small
\caption{List of emission lines used in the present study. Lines marked with asterisks (*) are density sensitive lines and used only to calculate electron number densities.}
\begin{tabular}{llc}
\hline\hline 
\textbf{Ion} & \textbf{Wavelength  (\AA)$^a$ } & \textbf{T$_{peak}$ (MK)$^a$} \\
\hline
Fe {\sc x} & 184.537 &      1.12          \\
Fe  {\sc xi} & 180.401, 182.167$^*$ &   1.37             \\
Si {\sc x} & 258.374$^*$, 261.056 &  1.41                  \\
S {\sc x} & 264.231 &  1.55             \\
Fe {\sc xii} & 192.394, 196.640$^*$ &   1.58             \\
Fe {\sc xiii} & 196.525$^*$, 202.044 &   1.78              \\
Fe {\sc xiv} & 264.789$^*$, 274.204$^b$ &   2.00              \\
Fe {\sc xv} & 284.163$^c$ &    2.24             \\
Fe {\sc xvi} & 262.976 &   2.82             \\
\hline
\end{tabular}
\\
\begin{flushleft}
$^a$ {Wavelengths and peak formation temperatures are taken from CHIANTI database.}\\
$^b$ {blended with Si  {\sc vii} 274.180 \AA .}\\
$^c$ {blended with Al {\sc ix} 284.042 \AA . }\\
\end{flushleft}
\label{tab:lines}
\end{table}

\begin{figure*}[htbp]
\centering
\includegraphics[width=0.85\textwidth]{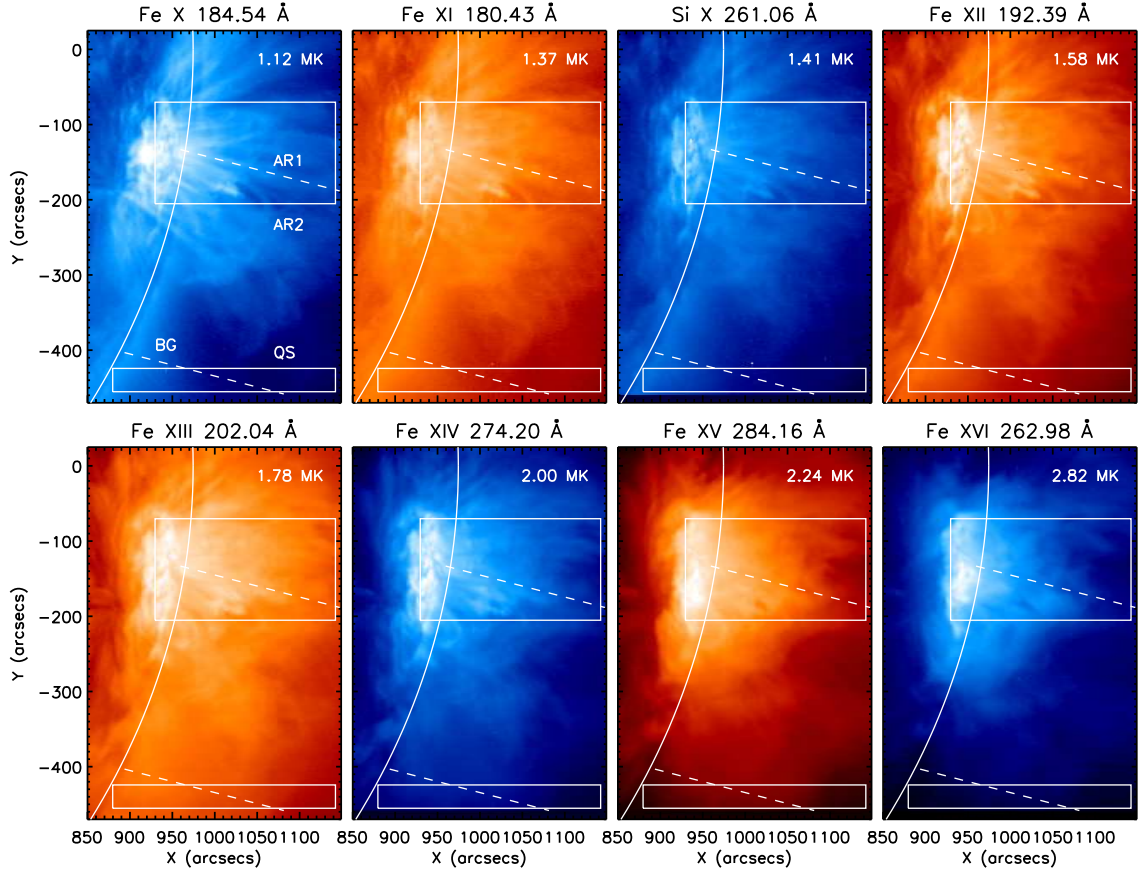}
\caption{Monochromatic intensity maps of off-limb active region and quiet Sun obtained in different wavelengths using EIS/Hinode (as labeled).
Dashed white lines  on each panel indicate off-limb locations (active region AR1, AR2, and quiet Sun QS) chosen for detailed analysis.
Continuous white  line on each panel indicates location of solar limb.}
\label{fig:context_loc}
\end{figure*}

In Figure~\ref{fig:context_loc}, we plot monochromatic intensity maps of off-limb active region obtained from different emission lines formed over temperature range
of 1.1 MK to 2.8 MK.  Intensity maps clearly show that
structures in the active region are not  well defined as discrete loop, instead emissions are more likely diffuse in nature without any sharp boundaries. Diffuse
emissions observed from active region in Fe~{\sc x}--Fe~{\sc xvi} lines are real. Such diffuse emissions were also highlighted by several authors in the past
\citep[e.g.,][]{2011A&A...525A.137O}. Here, we show diffuse nature of active region with several  emission lines formed over 1.1 MK to 2.8 MK.

To study variation of several physical parameters with height, we chose several structures and stripes in the off-limb active region and quiet Sun.
Although active region is diffuse in nature, we can still identify some bright loop-like structures extending far in to the corona from Fe~{\sc xii} 192.394 \AA\ intensity map. 
We traced and analyzed several such structures. Here, we present result from one such structure which was traced up to very
far off-limb distance. Traced stripe is named as AR1. Furthermore, to get the average behavior of active region, we binned over whole active region data in the 
solar-Y direction. Similarly, we also binned over small quiet Sun region in the solar-Y direction to study the quiet Sun.
Boxes chosen to obtain the average data are labeled as AR2 and QS. We also traced another stripe parallel to
AR1 in the quiet Sun region only for the background study purpose.  All the chosen stripes and boxes are shown in Figure~\ref{fig:context_loc}. 

\begin{figure*}[htbp]
\centering
\includegraphics[width=0.85\textwidth]{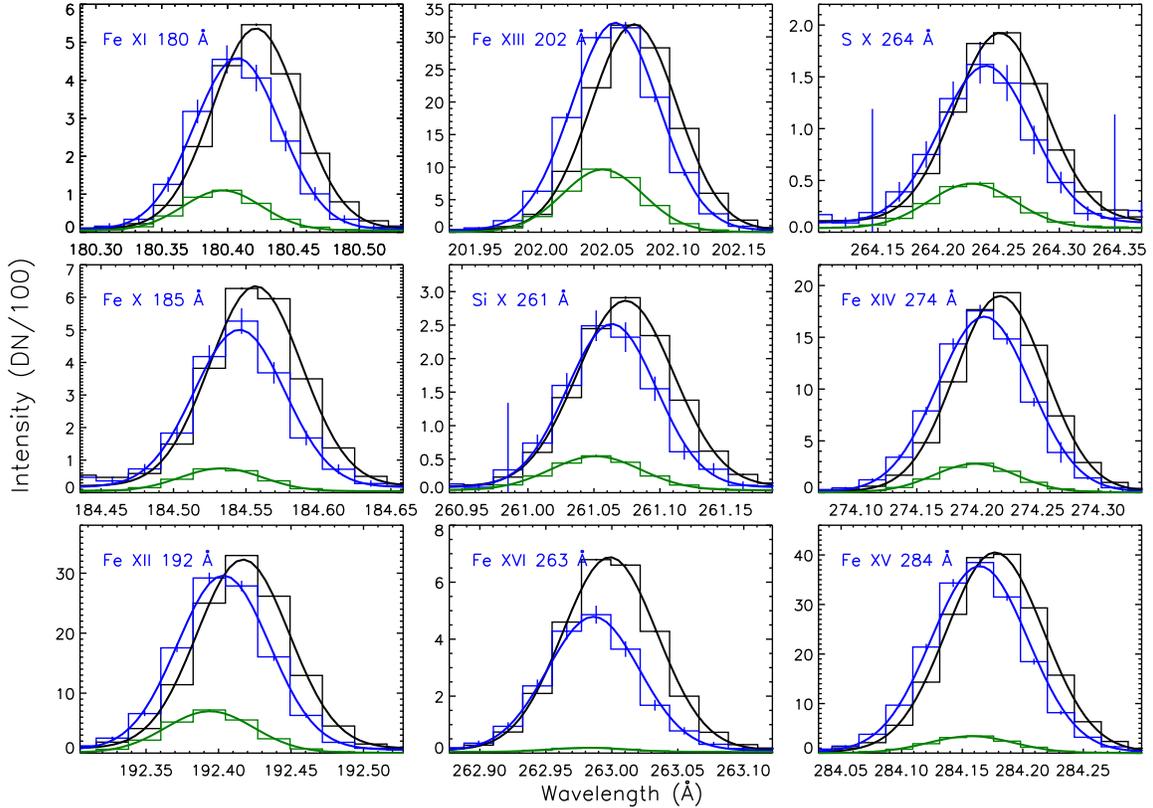}
\caption{Examples of spectral line profiles and fitted Gaussian profiles obtained at the off-limb height of 61 Mm along AR1 (blue lines), AR2 (black lines), and QS (green lines).}
\label{fig:profiles}
\end{figure*}

As selected spectral lines are isolated and unblended, we fitted all the profiles with single Gaussian function.
In Figure~\ref{fig:profiles}, we show examples of spectral line profiles and fitted Gaussian. Profiles were obtained at the 
off-limb distance of 61 Mm along AR1, AR2, and QS. From the plots, it is clear that all the profiles are symmetric and
can be well represented by a single Gaussian function.

One of the factor which may affect our analysis would be contamination from instrumental scattered or stray light.  
Minimum stray light contribution above the dark current is found to be around 2\% of the total on-disk counts for that
respective line\footnote{\url ftp://sohoftp.nascom.nasa.gov/solarsoft/hinode/eis/doc/eis\_notes/12\_STRAY\_LIGHT/eis\_swnote\_12.pdf}.
We chose sufficient box size to calculate average counts from each spectral line in the on-disk part of Sun. 
As stray light contamination in the off-limb corona will be simply 0.02 times of average counts, 
we obtained fraction of stray light contribution along all the stripes for all the lines.
As intensity drops-off with height in the off-limb corona, stray light contribution increases  with height.
We found stray light contributions to be  less than 8\%  up to the off-limb distance of $\approx 140$ Mm along AR1 and AR2
in all the spectral lines except for Fe~{\sc xvi} 263 \AA. These contributions were obtained after using  2\% weightage of on-disk counts.
For Fe~{\sc xvi} 263 \AA\ spectral line, stray light contributions were  below 9\% up to the distance
of 100 Mm. Beyond that distance, contribution increases sharply to about 20\% and 35\%  along  AR1 and AR2 respectively.
For the quiet Sun region QS, stray light contributions were less than 15\%  up to very far distance ($\approx 125$ Mm) in most of the spectral lines.
Stray light contributions in Fe {\sc x} 185 \AA , Fe~{\sc xiv} 274 \AA , and Fe~{\sc xv} 284 \AA\ lines increase sharply to above 40\% 
for distances beyond 130 Mm, whereas that in Fe~{\sc xvi} 262.976 \AA\ goes beyond 100\% for most of the distances along QS.
This may indicate that signal in Fe~{\sc xvi} 262.976 \AA\  line along QS is mainly from scattered light. Near the limb, stray light contributions 
for all the stripes are much smaller ($< 3\%$ for heights $< 80$ Mm and  $< 90$ Mm along AR1 and AR2 respectively, whereas
$< 5\%$  for heights $< 50$ Mm  for QS).  
As noted by \citet{2012ApJ...753...36H}, stray light contaminations start affecting line width measurements only if its
contributions are more than 45\%. In the current analysis, as stray light contributions are found to be very small, thus, it effects
are almost insignificant in the line width measurements. Results obtained in the present analysis are mainly derived from
heights where stray light contaminations are very small for most of the spectral lines ($< 8\%$ for distances up to 
$\approx$140 Mm along AR1 and AR2, and $<15\%$ for distances up to $\approx$125 Mm along QS).

\section{Results} \label{sec:results}

 As per  Equation~\ref{eq:energyf}, Alfv\'en wave energy flux is proportional to  $\sqrt{N_e} \xi ^2 BA$. 
We describe estimation of electron number density and wave velocity amplitude in the following subsections.

\subsection{Intensity, Density, and Emission Measure}\label{sec:intdens}

\begin{figure*}[htbp]
\centering
\includegraphics[width=0.8\textwidth]{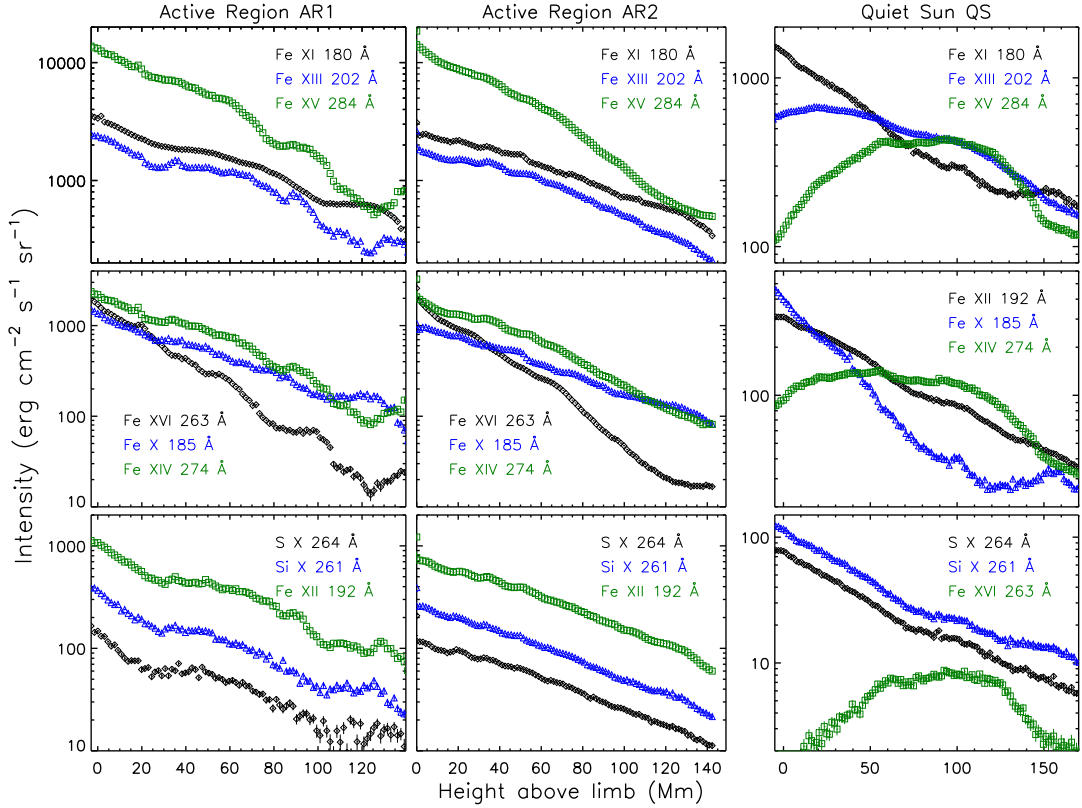}
\caption{Intensity variation with height along active region AR1, AR2, and quiet Sun QS obtained from various spectral lines as labeled.
True errors will also include  22\% uncertainty of the observed intensity based on the pre-flight calibration of EIS \citep{2006ApOpt..45.8689L}.}
\label{fig:int_fall}
\end{figure*}

 In Figure~\ref{fig:int_fall}, we plot variation of intensity obtained from selected spectral lines  with height along active region
 AR1, AR2, and quiet Sun QS. Associated $1 \sigma$ errorbars are also plotted with the data-points. 
From the plots, it is clear that dataset  has good signal strength in all the selected spectral lines  in the off-limb region of solar corona.
This makes it suitable to estimate electron number density for the calculation of  total Alfv\'en wave energy flux with height.

Electron number densities obtained from various spectral line pairs in the corona can be compared with hydrostatic equilibrium model.
This have been done in the past with imaging observations  using Transition Region and Coronal Explorer \citep[TRACE; e.g.,][]{1999ApJ...515..842A}, 
and recently with spectroscopic observations using EIS/Hinode \citep[e.g., ][]{2014ApJ...780..177L,2015ApJ...800..140G}. 

Electron number density profile in hydrostatic equilibrium is given by,

\begin{equation}
 N_e(h)=N_e(0)exp \left(- \frac{h}{\lambda (T_e)} \right)
\label{eq:hydrostatic}
\end{equation}

where $\lambda$ is density scale height given by,

\begin{equation}
 \lambda(T_e)= \frac{2k_b T_e}{\mu m_H g} \approx 46 \left[ \frac{T_e}{1~MK} \right] [Mm]
\label{eq:stemp}
\end{equation}

where $k_b$ is the Boltzmann constant, $T_e$ is electron temperature, $\mu$ is  mean molecular weight
 ($\approx$ 1.4 for the solar corona), $m_H$ is mass of the hydrogen atom, and $g$ is acceleration due
 to gravity at the solar surface \citep[see e.g.,][]{1999ApJ...515..842A}. 
 
 Moreover, observationally measured quantity such as intensity of an optically thin emission line depends on electron number density, i.e. 
 $I\propto N_e^\beta$ where $1< \beta< 3$, and  value of $\beta$ depends upon whether the given line is allowed, forbidden,
 or inter-system \citep{1994A&ARv...6..123M}. In this study, we have chosen only allowed lines to do the line width analysis.
 Some density sensitive forbidden lines were also chosen to calculate electron number density  (see Table~\ref{tab:lines}).
 In Figure~\ref{fig:int_fall}, we plot variation of intensity obtained from all the spectral lines with height along AR1, AR2, and QS stripes.

Since  dataset covers active and quiet Sun regions over wide range of wavelength, we identified several density sensitive line pairs formed over range of temperature.
We selected  Fe  {\sc xi} $ \lambda 182.167/ \lambda 180.401$, Si {\sc x} $ \lambda 258.374/ \lambda 261.056 $, Fe {\sc xii} $ \lambda 196.640/ \lambda 192.394$,  
Fe {\sc xiii} $ \lambda 196.525/ \lambda   202.044 $, and  Fe {\sc xiv} $ \lambda 264.789/ \lambda  274.204 $ line pairs 
to obtain the electron number density \citep{2007PASJ...59S.857Y}. In order to perform density and temperature diagnostics on the active region loops,
background subtraction plays an important role \citep[e.g.,][]{2003A&A...406.1089D}. \citet{2011A&A...525A.137O} analyzed the current dataset
before and used quiet Sun region to study background emission. We follow the same strategy and use intensity along the quiet Sun
stripes to perform background subtraction along active region stripes. Therefore, we use quiet Sun stripes BG and QS to subtract background emission
from active region stripes AR1 and AR2 respectively (stripes are shown in Figure~\ref{fig:context_loc}). 
All plasma diagnostics are performed over these background subtracted intensities.

\begin{figure*}[htbp]
\centering
\includegraphics[width=0.85\textwidth]{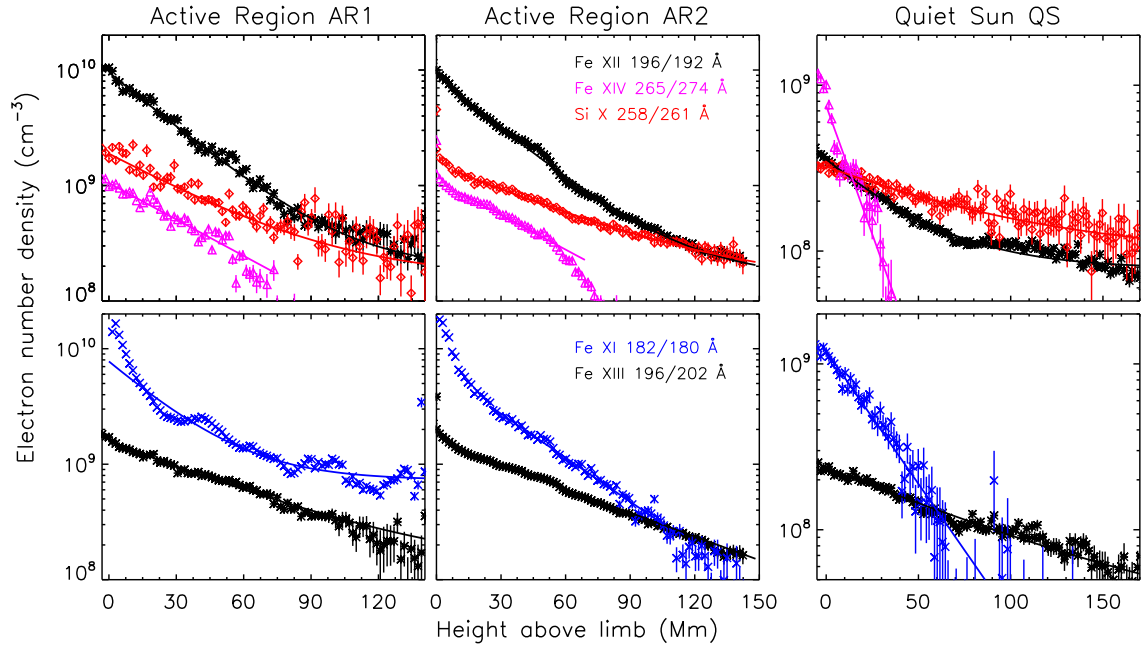}
\caption{Electron number density variation with height along active region AR1, AR2, and quiet Sun QS obtained from different density sensitive spectral line pairs  as labeled.
Over-plotted continuous lines represent fitted exponential decay profile to obtain density scale heights from various spectral line pairs (see also Table~\ref{tab:dens_scale}).}
\label{fig:dens_fall}
\end{figure*}

In Figure~\ref{fig:dens_fall}, we plot variation of electron number density derived from selected spectral line pairs  with height along  AR1, AR2, 
and QS. Plots show that as height increases electron number density decreases, however, corresponding errorbar increases with height.
Some of the lines in quiet Sun region show estimates with larger errorbars. Near the active region limb, electron number 
densities were estimated to be of the order of  $>10^9$ cm$^{-3}$ which drops to around $10^8$ cm$^{-3}$ in the far off-limb region. Densities
obtained from Fe {\sc xi}, and Fe {\sc xii} line pairs show almost similar numbers, whereas those obtained from Fe {\sc xiii}, and Si {\sc x} line pairs show similar values.
Near the limb region, densities from Fe {\sc xi}, and Fe {\sc xii} pairs show consistently larger values than that of Fe {\sc xiii}, and Si {\sc x} pairs. 
However, they all seem to converge towards similar values  beyond the distance of 80 Mm ($<5\times10^8$ cm$^{-3}$) and 95 Mm
($<4.5\times10^8$ cm$^{-3}$) along AR1 and AR2 respectively. Densities estimated from Fe {\sc xiv} line pair are smaller in number
as compared to other pairs and also falls-off more rapidly with height in both AR1 and AR2.  

In the quiet Sun region, we found number densities  to be lower than those in active region. In this case, densities estimated from Fe {\sc xii}, 
and Fe {\sc xiii} line pairs converge to similar numbers beyond 45 Mm ($<1.6\times10^8$ cm$^{-3}$). However, near the limb, densities obtained from 
Fe {\sc xii} pair are higher than that from Fe {\sc xiii} pair. Densities obtained from Si {\sc x} pair are higher than that estimated from
Fe {\sc xii}, and Fe {\sc xiii} pairs. Densities estimated from Fe  {\sc xi} and  Fe {\sc xiv} pairs near the limb are comparatively higher than those 
obtained from other line pairs but drops-off very rapidly with height.

We fitted electron number density variation with height along AR1, AR2, and QS  with exponential function
$N_e=N_0 exp(-h/H_d)+c$ using MPFIT routines \citep{2009ASPC..411..251M}. Fits provide density scale heights $H_d$ at different 
temperatures  obtained from different spectral line pairs (see Table~\ref{tab:dens_scale}). Expected electron density scale heights  $\lambda(T_e)$
as per hydrostatic equilibrium model at at different temperatures (see Equation~\ref{eq:stemp}) are also provided in the table. Comparison between
two density scale heights indicate that both the active and quiet Sun regions are basically underdense with few exceptions from quiet Sun region.  

\begin{table*}[htbp]
\centering
\small
\caption{Electron number density scale heights obtained from various spectral line pairs along active region AR1, AR2, and quiet Sun QS.}
\begin{tabular}{lcccccc} 
\hline\hline 
\textbf{Ion} & \textbf{Wavelength} & \textbf{T$_{peak}$} & \textbf{Hydrostatic}      & \multicolumn{3}{c}{\textbf{Density scale height}  (Mm)} \\
                  &    (\AA )          &           (MK)                          &  \textbf{height} (Mm)    & \textbf{AR1}  & \textbf{AR2}  & \textbf{QS} \\
\hline
Fe  {\sc xi}  & 182.167/180.401 &   1.37   & 63.02  & $ 24.86 \pm 0.31 $ & $ 34.22 \pm 0.26 $  & $ 27.13 \pm 0.98 $           \\
Si {\sc x}     & 258.374/261.056 &   1.41   & 64.98  & $ 39.84 \pm 2.15 $ & $ 53.61 \pm 2.25 $  & $ 78.12 \pm 9.91 $            \\
Fe {\sc xii}  & 196.640/192.394 &   1.58   & 72.91  & $ 26.98 \pm 0.36 $ & $  28.54 \pm 0.20 $  & $ 38.70 \pm 0.66 $       \\
Fe {\sc xiii} & 196.525/202.044 &   1.78   & 81.80 & $ 55.23 \pm 1.44  $ & $ 63.44 \pm 0.22 $  & $ 90.61 \pm 5.48  $      \\
Fe {\sc xiv} & 264.789/274.204 &   2.00   & 91.78 & $ 41.37 \pm 0.56 $ & $  43.53 \pm 0.32 $  & $ 14.00 \pm 0.96  $      \\
\hline
\end{tabular}
\label{tab:dens_scale}
\end{table*}

\begin{figure*}[htbp]
\centering
\includegraphics[width=0.8\textwidth]{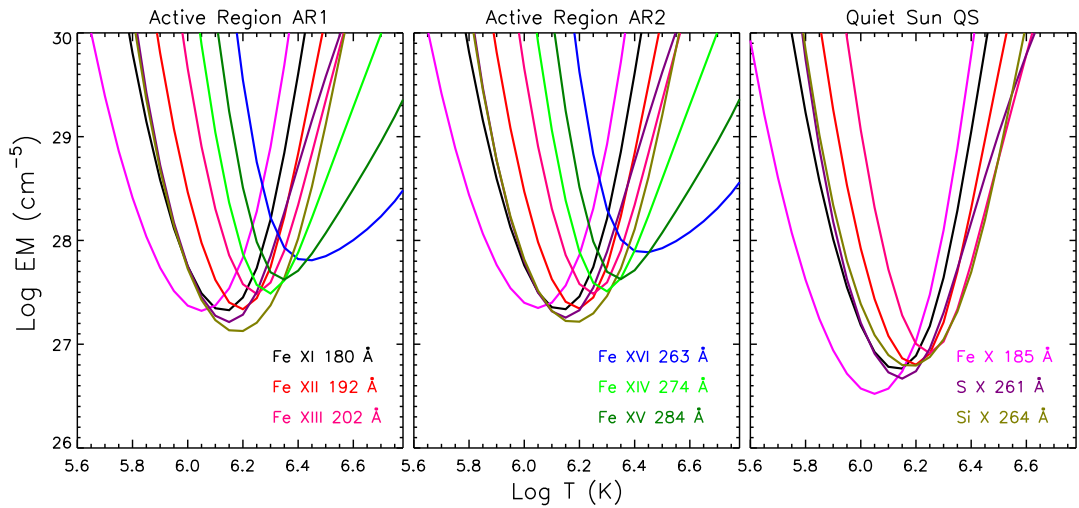}
\caption{EM loci plots obtained at off-limb height of 55 Mm along AR1, AR2, and QS.}
\label{fig:emlocii}
\end{figure*}

As active region and quiet Sun stripes were observed over a range of temperature, we employed emission measure (EM) loci technique
 to examine the thermal structure of  different stripes as a function of off-limb height.
 Several EM loci plots were constructed at different heights along the active region AR1, AR2, and quiet Sun QS stripes.
 In Figure~\ref{fig:emlocii}, we present sample EM loci plots obtained at the height of 55 Mm above the off-limb. 
From the plots, it is clear that plasma along the line-of-sight is not isothermal at that height.
Based on EM loci plots, distribution of plasma  along all the stripes were found to be multi-thermal at all the heights.
These results are in good agreement with the findings of  \citet{2011A&A...525A.137O} who analyzed the same dataset before.
They found  plasma in the active region to be multi-thermal at different distances from the limb.
Similarly, \citet{2008ApJ...686L.131W} also studied isolated coronal loops from the same active region when observed on-disk
and found them to be not isothermal. Thus, based on current and previous studies, plasma along the different active region
stripes can be considered as multi-thermal. This may indicate that emission in different lines are coming from either
single coronal structure formed over wide range of temperature or there exist multiple structures at different temperatures 
along the line-of-sight.  Moreover, plasma along the off-limb quiet Sun region appears to be nearly isothermal if 
the contributions from hot lines are excluded. Figure~\ref{fig:int_fall} shows that intensities obtained from hot lines along QS are increasing
with height near off-limb regions. This may suggest some possible contaminations in the hot lines from nearby active region.

\subsection{Non-thermal Velocity}
\label{sec:ntv}

\begin{figure*}[htbp]
\centering
\includegraphics[width=0.8\textwidth]{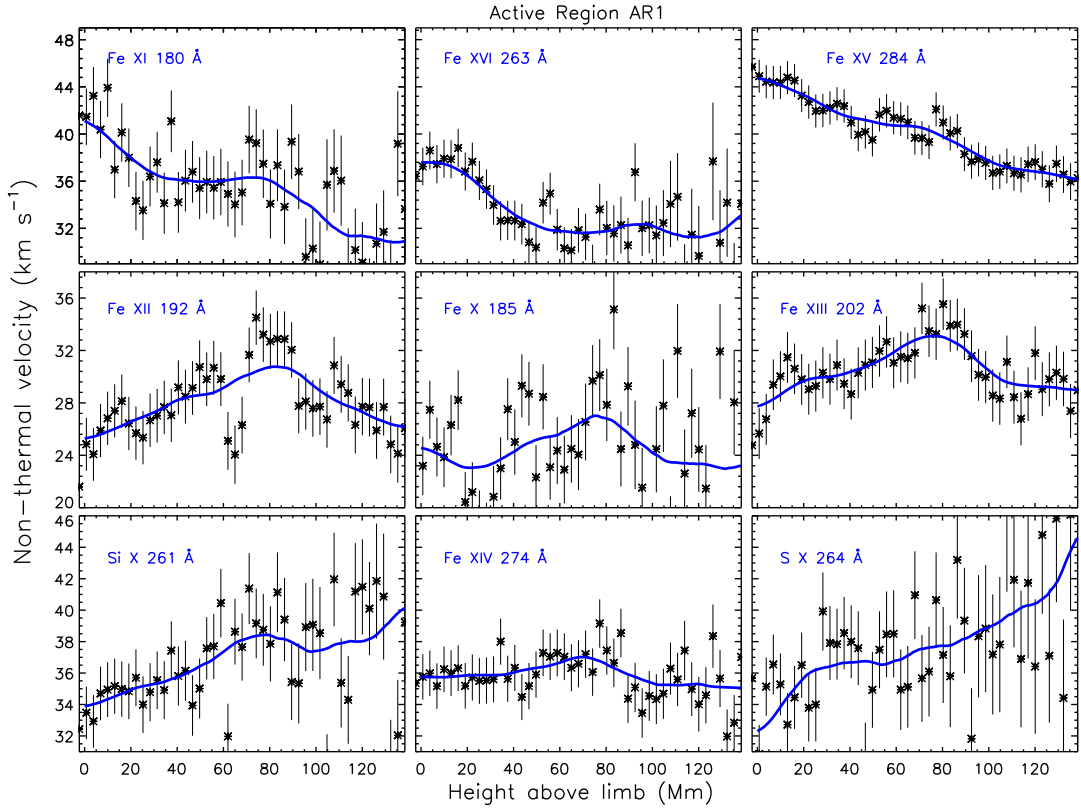}
\caption{Variation of non-thermal velocity with height along active region AR1 obtained from various spectral lines as labeled.
Overplotted continuous lines in all the panels show smooth variation of data-points obtained using 20-point running average. }
\label{fig:vnt_ar1}
\end{figure*}

\begin{figure*}[hbtp]
\centering
\includegraphics[width=0.8\textwidth]{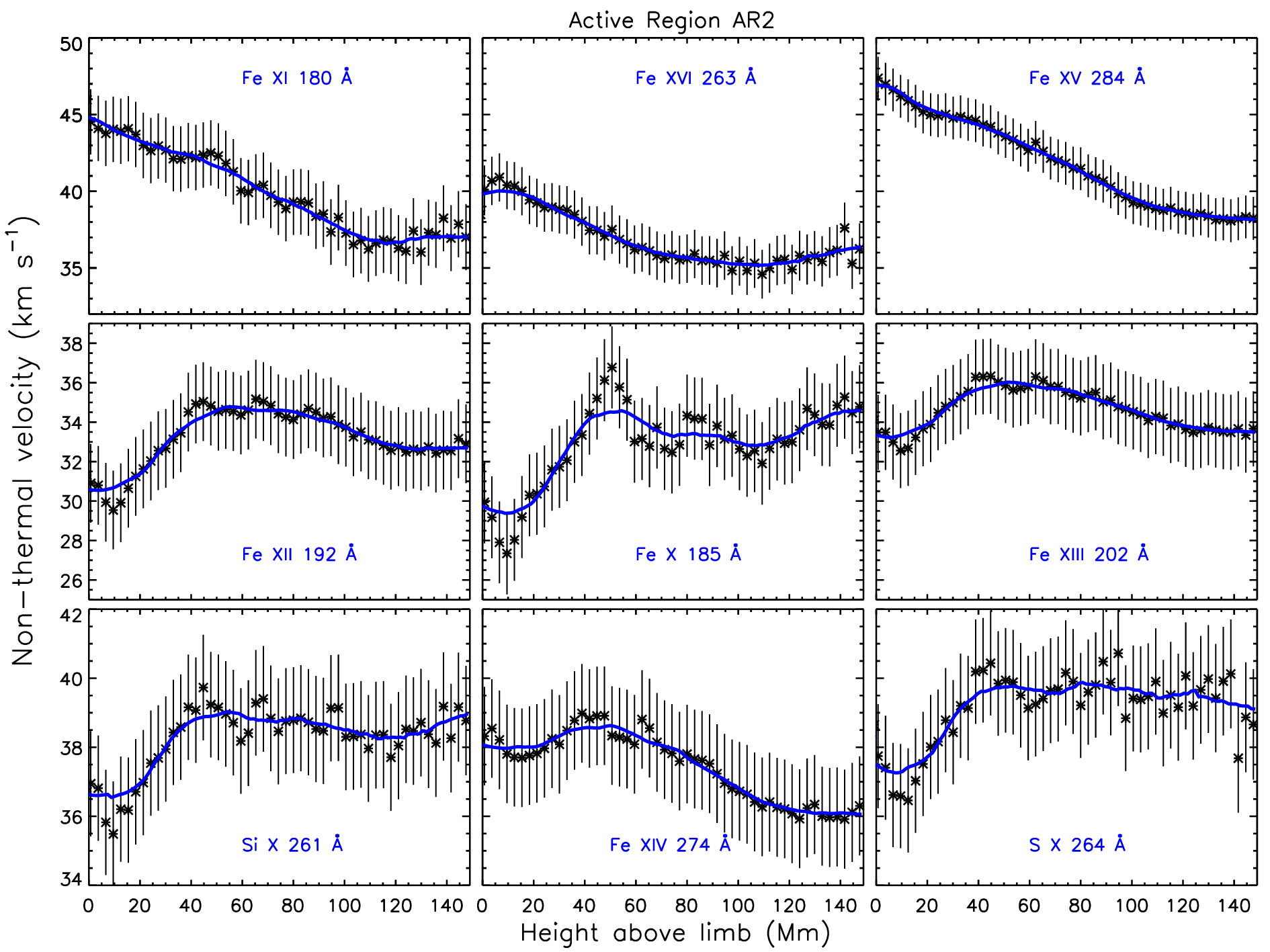}
\caption{Same as Figure~\ref{fig:vnt_ar1} but for active region AR2.}
\label{fig:vnt_ar2}
\end{figure*}

Non-thermal velocities are an important ingredient for calculation of Alfv\'en wave energy flux. These have been extracted from the observed 
emission line profiles as follows. Observed full width half maximum (FWHM) of any coronal spectral line is given by,

\begin{equation}
FWHM = \left[4 ln 2 \left(\frac{\lambda}{c}\right)^2 \left(\frac{2 k_b T_{i}}{M_{i}} +{\xi}^2\right) + W^{2}_{inst}\right]^{1/2}
\label{eq:fwhm}
\end{equation}

where $T_{i}$ is ion temperature, $M_{i}$ ion mass,  $\xi$ is non-thermal velocity, and $W_{inst}$ is the instrumental width. 
EIS/Hinode instrumental width is not constant, and is found to vary with CCD Y-pixel position along the 
slit\footnote{\url ftp://sohoftp.nascom.nasa.gov/solarsoft/hinode/eis/doc/eis\_notes/07\_LINE\_WIDTH/eis\_swnote\_07.pdf}.   
EIS instrumental width for $2\arcsec$ slit varies between 64--74 m\AA\ for a downloaded central 512-pixels
(starting from pixels 256 to 767). These widths were calculated using IDL routine  EIS\_SLIT\_WIDTH provided by EIS team.
Instrumental widths were then subtracted from the FWHM of spectral lines accordingly.
We further calculated non-thermal components by subtracting thermal components from each spectral lines. Thermal components were 
calculated after assuming ion temperatures to be equal to the peak formation temperature of spectral lines as found from contribution functions (see 
Figure~\ref{fig:cont_func} and Table~\ref{tab:lines}). After subtraction of instrumental width, line widths were  primarily dominated 
by non-thermal components. Errorbars on non-thermal velocities were calculated using errors in the profile fitting, 3 m\AA\ error in instrumental width,
and errors in assumed thermal temperatures which were taken to be half width half maxima (HWHM) of Gaussian fits applied to contribution functions
of respective spectral lines. 

\begin{figure*}[htbp]
\centering
\includegraphics[width=0.8\textwidth]{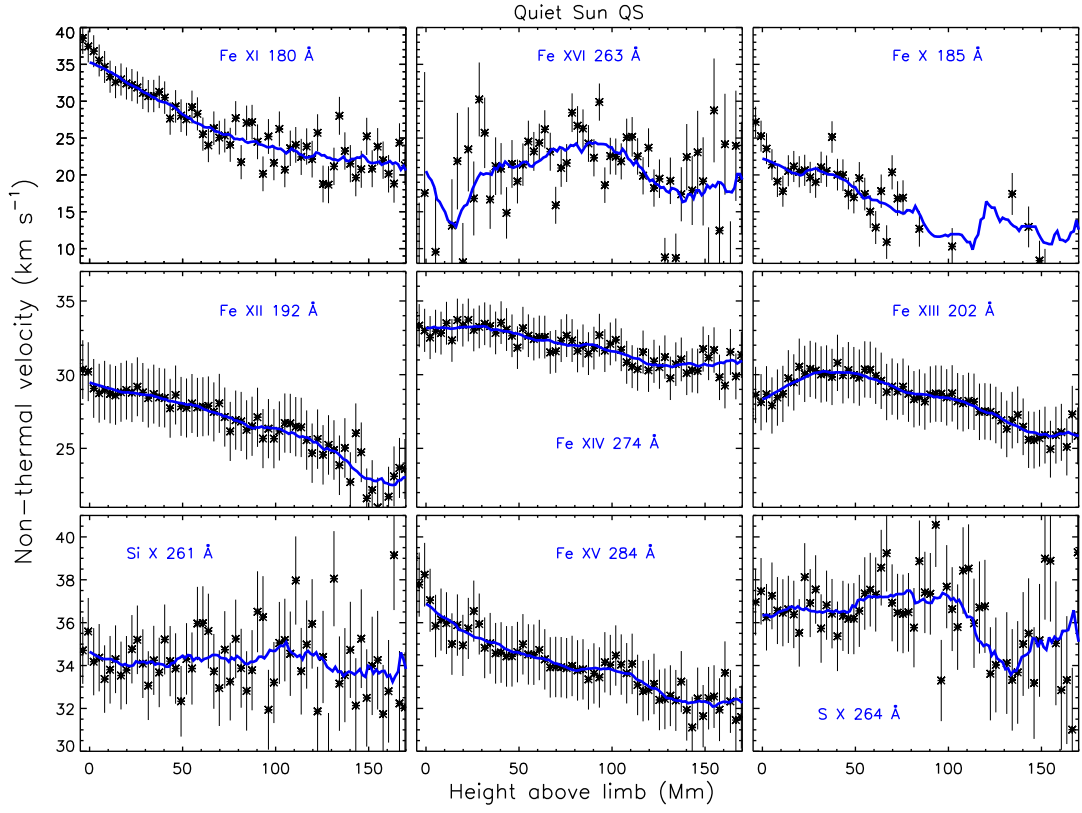}
\caption{Same as Figure~\ref{fig:vnt_ar1} but for quiet Sun QS.}
\label{fig:vnt_qs}
\end{figure*}

In Figures~\ref{fig:vnt_ar1}, \ref{fig:vnt_ar2}, and \ref{fig:vnt_qs}, we plot non-thermal velocities ($\xi$) obtained from various spectral lines with height
along active regions AR1, AR2, and quiet Sun QS respectively. We also over-plot 20-point running average of data-points to visualize variations on longer spatial scale.
Non-thermal velocities obtained from warm lines such as Fe~{\sc x} 185 \AA , Fe~{\sc xii} 192 \AA , and Fe~{\sc xiii} 202 \AA\
show initial increase from $\approx 24$ km s$^{-1}$ near the limb to  $\approx 33$ km s$^{-1}$ at around height of 80 Mm, whereas those
obtained from Si~{\sc x} 261 \AA , and S~{\sc x} 264 \AA\ show increase from $\approx 34$ km s$^{-1}$ to $\approx 39$ km s$^{-1}$
at the similar heights along AR1. Beyond these heights, non-thermal velocities either decrease or remain almost constant with some scattered data-points.

Variation of non-thermal velocities with height along AR2 also show similar pattern as in AR1 but their values are enhanced by   $\approx 2-3$ km s$^{-1}$.
This possible enhancement along AR2 could  be due to  integration taken over larger spatial scale to deduce the  non-thermal velocities.
On comparison with polar regions  \citep[e.g.,][]{2009A&A...501L..15B,2012ApJ...751..110B},  non-thermal velocities obtained from
 Fe~{\sc xii} line in the active regions are consistently smaller in magnitude but shows sharp increase with height. However, recent findings of 
 \citet{2014ApJ...780..177L} showed consistent decrease in non-thermal velocities along the cool loop and dark lane in the off-limb active region.
Moreover, non-thermal velocities obtained from hot lines such as Fe~{\sc xv} 284 \AA , and Fe~{\sc xvi} 263 \AA\ show gradual decrease with height.
Velocities obtained from hot Fe~{\sc xv} 284 \AA\ line shows decrease from $\approx 45$ km s$^{-1}$ near the limb to $\approx 36$ km s$^{-1}$ beyond 100 Mm
 along AR1, whereas Fe~{\sc xvi} 263 \AA\ line shows decrease from $\approx 38$ km s$^{-1}$ to $\approx 32$ km s$^{-1}$.
Variations obtained from hot Fe~{\sc xv} 284 \AA , and Fe~{\sc xvi} 263 \AA\ lines along AR2 show pattern again similar to that in AR1 with
velocities being again enhanced by $\approx 2-3$ km s$^{-1}$. Surprisingly, warm Fe~{\sc xi} 180 \AA\ line shows pattern similar to that of hot lines
whereas Fe~{\sc xiv} 274 \AA\ line shows intermediate behavior of hot and warm lines. 
\citet{2006ApJ...639..475S} performed line width study along steady coronal structures using data from Norikura coronagraph.
They found decrease in FWHM of hot Fe~{\sc xiv} 5303 \AA\ line up to the distance of $300''$ above the limb which became constant thereafter. 
They also found increase in FWHM of warm Fe~{\sc x} 6374 \AA\ line up to the distance of $250''$ which remained unchanged further.
FWHM of intermediate lines (Fe~{\sc xi} 7892 \AA\ and Fe~{\sc xiii} 10747 \AA ) showed intermediate behavior. Thus, results of
line width variation with height do indicate some temperature dependence. Findings in this study are almost similar to those
of  \citet{2006ApJ...639..475S} with some shift in temperature dependence. This shift might be specific to the active regions
studied. However, cause for exceptional behavior of warm Fe~{\sc xi} 180 \AA\ line in this study is unknown and can not be 
speculated at this stage. Recently, \citet{2016ApJ...820...63B} surveyed 15 non-flaring on-disk active regions using EIS/Hinode. 
They measured non-thermal velocities at specific locations in the cores of solar active regions over the temperature range of 1--4 MK.
However, they did not find any significant trend with temperature.

In the quiet Sun region, non-thermal velocities obtained from warm iron lines show consistent decrease with height. Non-thermal velocities
obtained from warm Si~{\sc x} and S~{\sc x} lines show almost  constant value of $\approx 34$ km s$^{-1}$  and  $\approx 36$ km s$^{-1}$ respectively
with height, however, they do show some large scatter around.  Non-thermal velocities obtained from hot Fe~{\sc xiv} and Fe~{\sc xv} lines also show decrease with height
similar to warm lines. No visible pattern can be inferred from hot Fe~{\sc xvi} line as signal in this line in the quiet Sun region is mostly due to scattered light as mentioned earlier.
These finding are similar to those of \citet{2002A&A...392..319H} where they studied spectral line profiles of warm Mg~{\sc x} 625 \AA\ line from quiet clean corona. 
They found narrowing of emission lines as a function of height similar to findings in the current study. They attributed narrowing of profiles with height
to dissipation of wave activity.

\subsection{Alfv\'en wave energy flux}
\label{sec:energy}

\begin{figure*}[htbp]
\centering
\includegraphics[width=0.8\textwidth]{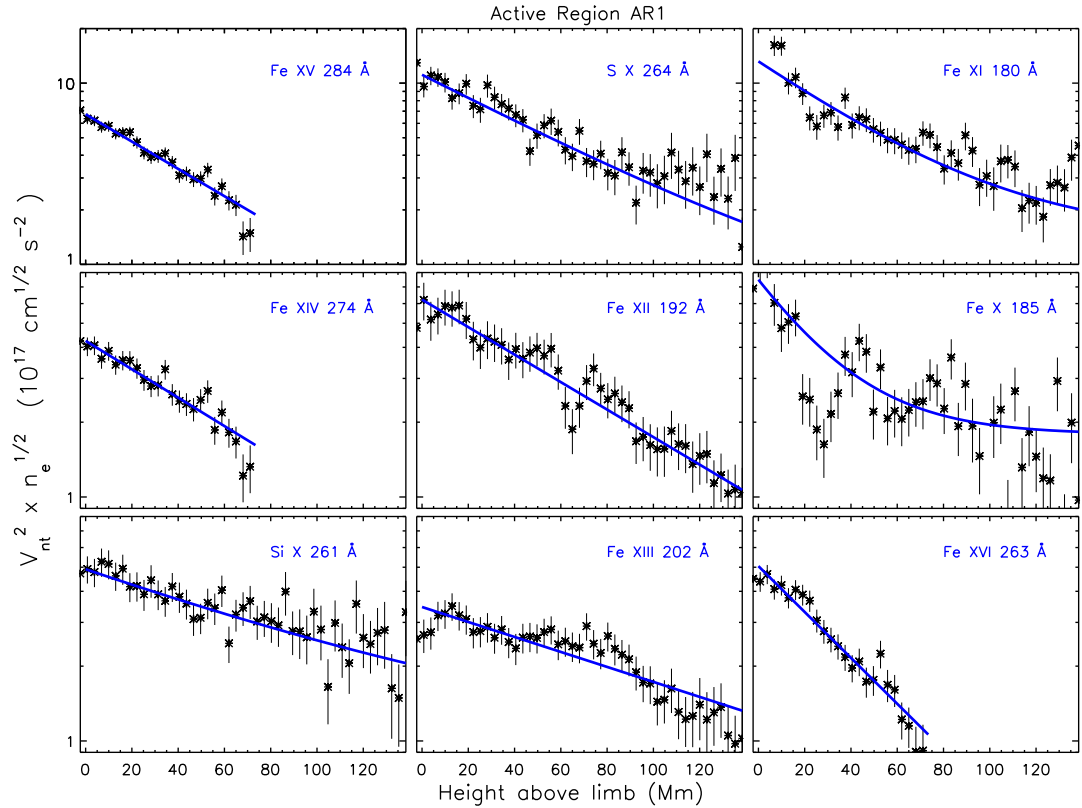}
\caption{Variation of proportional Alfv\'en wave energy flux ($\sqrt{N_e}~\xi^2$) with height along active region AR1 obtained from various spectral lines as labeled. Over-plotted continuous
lines are fitted exponential decay profile to obtain wave damping lengths from various spectral lines formed at range of temperature (see also Table~\ref{tab:damp_scale}).}
\label{fig:enrg_ar1}
\end{figure*}

\begin{figure*}[htbp]
\centering
\includegraphics[width=0.8\textwidth]{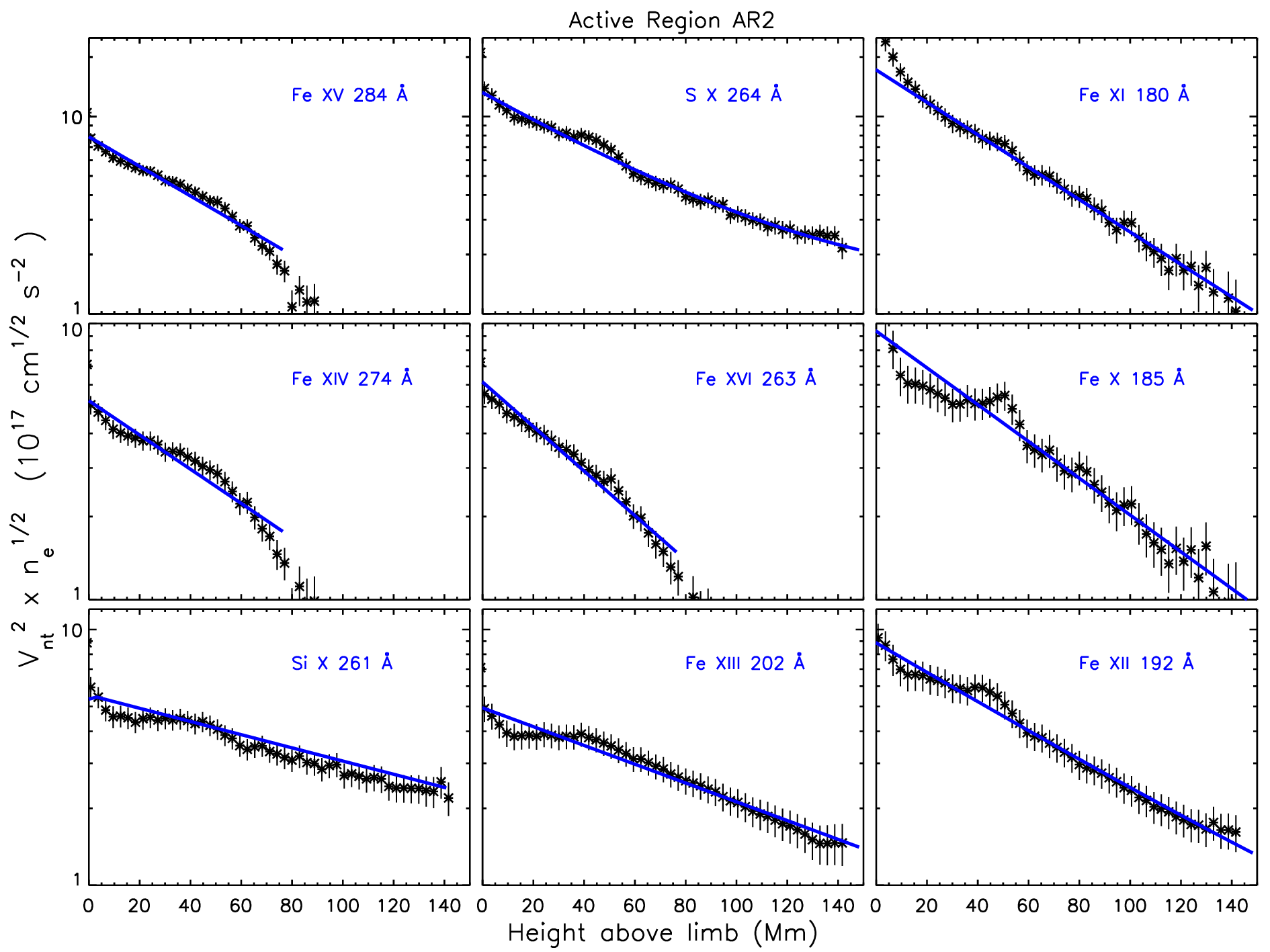}
\caption{Same as Figure~\ref{fig:enrg_ar1} but for active region AR2.}
\label{fig:enrg_ar2}
\end{figure*}

\begin{figure*}[htbp]
\centering
\includegraphics[width=0.8\textwidth]{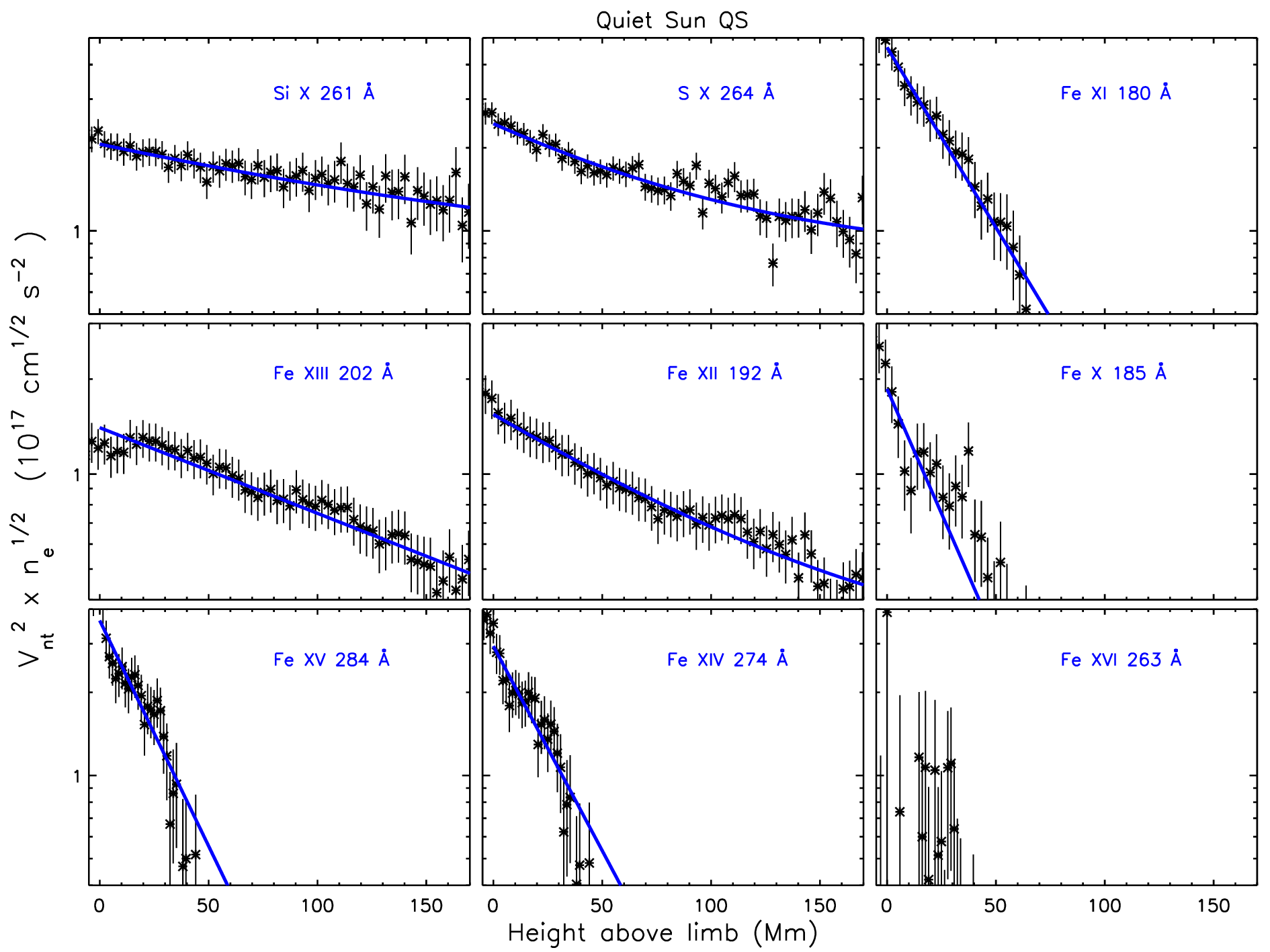}
\caption{Same as Figure~\ref{fig:enrg_ar1} but for quiet Sun QS.}
\label{fig:enrg_qs}
\end{figure*}

\begin{table*}[thbp]
\centering
\small
\caption{Damping lengths derived from various spectral lines along active region AR1, AR2, and quiet Sun QS.}
\begin{tabular}{lccccc} 
\hline\hline 
\textbf{Ion} & \textbf{Wavelength} & \textbf{ T$_{peak}$}   & \multicolumn{3}{c}{\textbf{Damping length $D_l$}  (Mm)} \\
                  &    (\AA )          &           (MK)                            & \textbf{AR1}  & \textbf{AR2}  & \textbf{QS} \\
\hline
Fe {\sc x}    & 184.537 &   1.12   &  $  28.31  \pm 6.72   $    &   $  65.19  \pm 3.45   $    &   $ 27.54  \pm 2.39  $   \\
Fe  {\sc xi}  & 180.401 &   1.37   & $   46.52  \pm 7.62  $    &   $   52.93 \pm  0.82  $    &   $  33.29  \pm 2.44   $   \\
Si {\sc x}     & 261.056 &   1.41   & $   114.87 \pm 77.31   $    &   $ 161.69   \pm 11.01   $    &  $  169.95 \pm  181.20 $   \\
S {\sc x}      & 264.231 &   1.55   & $  65.02  \pm 12.47  $    &   $  56.32   \pm 5.72  $    &   $ 84.47  \pm  23.77  $   \\
Fe {\sc xii}  & 192.394 &   1.58   & $  78.49 \pm  4.38  $    &   $  73.70  \pm 13.44  $    &   $  91.95  \pm 20.12 $   \\
Fe {\sc xiii} & 202.044 &   1.78   & $   144.1 \pm 12.44  $    &   $  118.30  \pm 6.85   $    &   $  159.98  \pm 10.49    $   \\
Fe {\sc xiv} & 274.204 &   2.00   & $  75.99 \pm  6.00  $    &   $ 70.06  \pm 4.45 $    &   $  29.31  \pm  3.50 $   \\
Fe {\sc xv}  & 284.163 &   2.24   & $  57.68 \pm  2.97 $    &   $ 57.99  \pm 2.66 $    &   $  26.59 \pm 2.79  $   \\
Fe {\sc xvi} & 262.976 &   2.82   & $  47.05 \pm  2.65   $    &   $ 53.85  \pm 2.90   $    &   $  ----  $   \\
\hline
\end{tabular}
\label{tab:damp_scale}
\end{table*}

Alfv\'en wave energy flux  can be calculated by using Equation~\ref{eq:energyf}. 
In a flux tube geometry, $B\times A$ will always be a constant. Because in a constant magnetic field model,
cross-sectional area will remain constant, thus, product will also remain constant. However, in the case of expanding flux 
tube model,  $B$ will decrease with height (let us assume inverse square field dependence),  whereas $A$ will increase with squared
radius dependence, thus, product of $B$ and $A$ will again be constant \citep[see ][]{2001A&A...374L...9M}.
Therefore, total Alfv\'en wave energy flux will always be proportional to $\sqrt{N_e}~\xi^2 $ in either case. Henceforth, if total Alfv\'en wave energy flux is conserved as
waves propagates outward, $\sqrt{N_e}~\xi^2 $ will remain constant with height. In Figures~\ref{fig:enrg_ar1}, \ref{fig:enrg_ar2}, and \ref{fig:enrg_qs},
we plot variations of $\sqrt{N_e}~\xi^2$ with height obtained from selected spectral lines along active region AR1, AR2, and quiet Sun QS respectively.
As electron number densities were estimated only from few spectral line pairs, therefore for rest of the lines, we choose number densities obtained
from line pairs formed at nearest temperature. Plots clearly show that product $\sqrt{N_e}~\xi^2$  decreases with height in all spectral lines
in all the regions. This provides clear evidence of damping of Alfv\'en wave energy flux with height in the both off-limb active and quiet Sun region.
Alfv\'en wave energy fluxes are found to be $\approx 1.85 \times 10^7 $ erg cm$^{-2}$ s$^{-1}$ near the limb which decreases to $\approx 0.86 \times 10^7 $
erg cm$^{-2}$ s$^{-1}$ at  around height of 70 Mm as calculated from Fe~{\sc xii} 192 \AA\ spectral line. To calculate the Alfv\'en wave energy flux, 
we assumed coronal magnetic field strength of 39 G as measured by \citet{2008A&A...487L..17V} using loop oscillations. 
Calculated Alfv\'en wave energy fluxes are of similar order of magnitude which is required to  maintain the active region corona
\citep[$\approx 10^7$ erg cm$^{-2}$ s$^{-1}$ as estimated by][]{1977ARA&A..15..363W}. 
Moreover, coronal magnetic field strength can vary like 10 G and 33 G as measured by  \citet{2000ApJ...541L..83L}  in two active regions at
distances of 0.12 and 0.15 $R_{\odot}$ using longitudinal Zeeman effect in Fe~{\sc xiii} 10747 \AA\ spectral line. Therefore, if assumed
 magnetic field strength is of the order of 10 G, then Alfv\'en wave energy fluxes will be slightly less than the energy flux required to maintain the corona.
One thing to be noted here that although Alfv\'en waves are getting gradually damped with height, non-thermal velocities obtained from warm spectral lines
were  initially increasing with height in the active region. This indicates that damping of Alfv\'en waves can only be inferred from complete calculation of
total Alfv\'en wave energy flux with height. Only non-thermal velocity estimates with height will not serve the purpose.

Upon finding the evidence of  damping of Alfv\'en wave energy flux with height, we further obtain damping length in all the spectral lines
covering range of temperature. Effect of damping can be calculated by multiplying $e^{-h/D_l}$ to the proportional  Alfv\'en wave energy flux
$\sqrt{N_e}~\xi^2 $, where $D_l$  is termed as \textquoteleft damping length\textquoteright\ for total Alfv\'en wave energy flux, 

\begin{equation}
F_{wt}\propto \sqrt{N_e}~\xi^2 e^{-h/D_l}
\label{eq:pdamp}
\end{equation}

\begin{equation}
F_{wt}\approx A~\sqrt{N_e}~\xi^2 e^{-h/D_l} +B
\label{eq:damping}
\end{equation}

where A and B are appropriate constants. Henceforth, we obtained damping length by fitting the  $F_{wt}$ values in different
spectral lines  as per  Equation~\ref{eq:damping}  using MPFIT routines \citep{2009ASPC..411..251M}.
Derived damping lengths $D_l$  from various spectral lines along active region AR1, AR2, and quiet Sun QS  are  in the range of 
25-170 Mm and provided in the Table~\ref{tab:damp_scale}.
 \citet{2012ApJ...751..110B} also reported decay of Alfv\'en wave energy flux with height in the polar coronal hole region.  However, they performed  linear fit to 
  decay profile and estimated decay rates to be $-1.07 \times 10^{-3} $ erg~cm$^{-1}$ below 0.03 R$_\odot$ and  $-4.5 \times 10^{-5} $ erg~cm$^{-1}$
 between 0.03-0.4 R$_\odot$. Equivalent damping length for the decay rate between 0.03-0.4 R$_\odot$ is calculated to be around 95 Mm. 
 They performed measurements using EIS Fe~{\sc xii} 195 \AA\ spectral line. In this work, damping lengths obtained from Fe~{\sc xii} 192 \AA\ lines are
 in the range of 75--90 Mm in both active and quiet Sun regions. This suggests that damping lengths obtained from both the studies are comparable.
 
\section{Discussions} \label{sec:discussions}

\begin{figure*}[hbtp]
\centering
\includegraphics[width=0.8\textwidth]{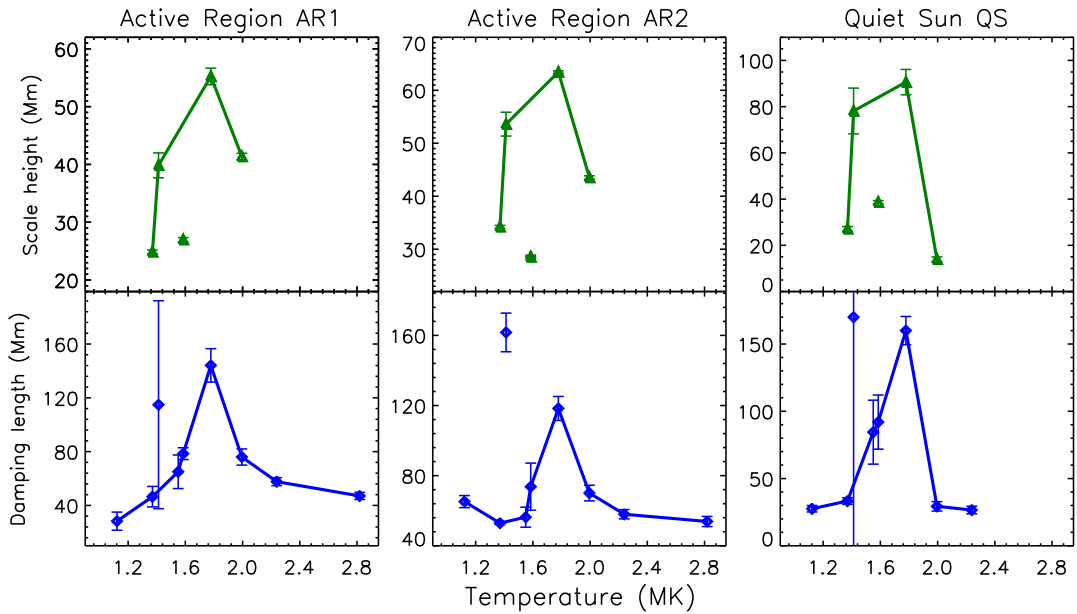}
\caption{Variation of density scale heights (top panels) and Alfv\'en waves damping lengths (bottom panels) obtained from different spectral lines with 
respect to their peak formation temperature.}
\label{fig:temp_damp}
\end{figure*}

In this work, we found clear evidence of damping of Alfv\'en waves in the off-limb active and quiet Sun region.
Damping lengths were found to be different for different spectral lines formed at different temperatures (see Table~\ref{tab:damp_scale}).
We further explore existence of any temperature dependence on various decay lengths obtained in this study.
Henceforth, we analyze density scale heights and Alfv\'en wave damping lengths  with peak formation temperature of their respective spectral lines (see Table~\ref{tab:lines}).
 We  plot density scale heights obtained from the different  line pairs with respect to their peak formation
 temperature (see top panels of Figure~\ref{fig:temp_damp}). Density scale heights  first  increase  and later decrease with 
 temperature. However,  density scale height obtained from Fe~{\sc xii} pair does not follow this trend.
  Moreover, hydrostatic scale heights as expected from  Equation~\ref{eq:stemp}
  are also provided in Table~\ref{tab:dens_scale}.  As mentioned earlier, comparison between the two density scale heights indicate that both active and 
 quiet Sun regions are basically underdense, with few exceptions from quiet Sun region. However, it appears that emissions coming from spectral lines
 formed near the temperature of 1.8 MK are more  closer to be in hydrostatic equilibrium than those formed different from 1.8 MK in the active region.
 This may be speculated as observed region to be filled with plasma of temperature nearly 1.8 MK and has poor supply for other cooler and hotter plasma.
 This result might be a characteristic of observed active region, and different active regions might have different temperature distribution.
 
In the bottom panels of Figure~\ref{fig:temp_damp}, we plot damping lengths obtained from
different spectral lines with respect to their peak formation temperature.  Different panels show that damping lengths first increase  and later decrease with 
temperature. Maximum damping length is attained at around temperature of 1.78 MK (corresponds to Fe {\sc xiii} 202 \AA) for all the active region and quiet Sun stripes. 
 We would also like to point out that several structures were traced and analyzed as mentioned earlier. Although obtained decay lengths were not same,
 results followed similar pattern for all the analyzed structures. Henceforth, obtained results indicate measurement of different damping lengths for different 
 temperatures. These results can either be interpreted as temperature dependent damping of Alfv\'en waves or measurement of different damping lengths in different 
 coronal  structures formed over wide range of temperature along the our line-of-sight.
 
Possible temperature dependent damping length of Alfv\'en waves may indicate that thermal conduction plays
some important role in the damping of these waves. However, role of thermal conduction in the damping of Alfv\'en waves is not much explored \citep[e.g.,][]{2011ApJ...736....3V}.
Although it is very well studied for the case of slow magneto-acoustic waves \citep[e.g.,][]{2002SoPh..209...89D}. Role of thermal conduction in the damping of slow magneto-acoustic
waves were recently observed by \citet{2014A&A...568A..96G} and \citet{2014ApJ...789..118K} based on possible period (of waves) dependent damping length.
In this study,  although we do not have any information on wave period, we have coverage over wide range of temperature. Work of \citet{2002SoPh..209...89D}
suggested that slightly enhanced thermal conductivity may explain observed damping lengths of 40-50 Mm for slow waves. These enhancements in thermal conductivity were
later also suggested by \citet{2014A&A...568A..96G}. In this study, observed damping lengths for Alfv\'en waves are in the range of 25-170 Mm as obtained from
different temperature lines. Henceforth, these results demand for  detailed investigation of role of thermal conduction in the damping of Alfv\'en waves.

Slow magneto-acoustic waves in the solar corona propagate along the field lines with propagation speed of the order of 100 km s$^{-1}$ and velocity amplitude of
the order of 5-10 km s$^{-1}$. Active region studied in this work is located near the limb and derived results are mainly focused on off-limb regions.
 In the off-limb region, magnetic field lines are generally found to be oriented nearly perpendicular to the observers line-of-sight. Therefore, contribution from
 observed Doppler velocities due to propagation of slow  magneto-acoustic waves in the measurement of non-thermal velocities will be minimal. 
 Similarly, studies on measurement of plasma flows in the active region loops indicate temperature dependent flow speeds.
 \citet{2008A&A...481L..49D} and \citet{2009ApJ...694.1256T} measured absolute flow speeds to be less than 30 km s$^{-1}$  along the active region loops
  using similar spectral lines  formed over temperature range of 0.6--2 MK. They found decrease in flow speeds 
   with increase in temperature (redshift to blueshift).  Moreover, \citet{2011ApJ...727L..13B} also measured on
  an average Doppler velocity of $-22$ km s$^{-1}$ from the edges of active regions.  
  Generally loops  cross an active region in the East-West direction, so flows along
  the off-limb loops will  either be directed toward or away from the observers line-of-sight (if loops are not radially directed). This may
  lead to some enhancements in the line-width.  However, as velocities in line-width measurements add in quadrature, contribution of Doppler velocities
 due to plasma upflows  ($<10$ km s$^{-1}$, due to inclination of loops along the line-of-sight) will again be minimal in the non-thermal velocities. 
 Moreover, there might be some enhancement in the non-thermal broadening
 due to these factors but given the range of errorbars (2--4 km s$^{-1}$), their  contribution can not be quantified.
 Measurements on AR2 which were obtained after taking average over larger spatial length,
 shows enhancements in non-thermal velocities by  $\approx 2-3$ km s$^{-1}$ as compared to measurements on AR1. 
 This could possibly be the effect of different Doppler shifted flows present along several different loop structures
 which were summed together to obtain the integrated profile, and thus, resulted in larger non-thermal velocities. 
 Henceforth, measured non-thermal velocities along AR2 can only be considered as an upper limit.
 
As mentioned earlier, role of  thermal conduction in the damping of slow magneto-acoustic waves is well known. One of the
possibility of  damping of Alfv\'en waves  would be that Alfv\'en wave energy is being transferred to slow magneto-acoustic waves.
These slow waves will further get easily dissipated  via thermal conduction and will finally show up as temperature dependent damping of Alfv\'en waves.
 \citet{2006A&A...456L..13Z} studied wave energy conversion process in  the non-linear ideal MHD framework. They demonstrated that wave energy 
 can be converted from Alfv\'en waves to slow magneto-acoustic waves near the region of corona  where plasma-$\beta$ approaches unity. 
As contributions from slow  magneto-acoustic waves in the current measurements of non-thermal velocities are minimal, henceforth, only
the damping of Alfv\'en waves can be inferred from the observed non-thermal velocities.
 
\section{Summary and Conclusions} \label{sec:conclusion}

We investigated off-limb active and quiet Sun region using spectroscopic data from EIS/Hinode. We studied height dependence of basic plasma
parameters such as intensity, electron number density, and non-thermal velocity along the active region and quiet Sun. These estimated 
parameters enabled us to further study height dependence of Alfv\'en wave  energy flux in the both regions. 
Main findings of our analysis are summarized as,

\begin{itemize}

 \item We identified several isolated spectral lines with good signal to noise ratio in the off-limb regions. These lines 
 are formed at different temperatures and cover the temperature range of 1.1--2.8 MK.
 
 \item We obtained electron densities and corresponding scale heights from different spectral line pairs which suggested that observed active
 and quiet Sun regions are basically underdense with few measured exceptions from quiet Sun region.
 
 \item Non-thermal velocities measured from warm spectral lines first showed increase with height and later showed either decrease or almost constant value with
 height in the far off-limb active region whereas hot lines showed  gradual decrease with height. However, those measured from various spectral lines in the quiet Sun
 region showed either gradual decrease or almost constant value with height.
 
 \item Calculated Alfv\'en wave energy fluxes were similar to or slightly less than the energy required to maintain the active region corona.
  Results also showed damping of Alfv\'en wave energy flux with height.
 
 \item We found damping lengths of Alfv\'en wave energy flux ($D_l$) to be in the range of 25-170 Mm as measured from different spectral lines
 formed at different temperatures.
 
 \item Variation of damping lengths first showed increase and later decrease with increasing temperature.
  Damping length peaked at around temperature of 1.78 MK in the both active and quiet Sun regions.
 
\end{itemize}

This work provides measurements of non-thermal velocities and Alfv\'en wave energy fluxes at wide range of temperature. Possible
interpretation of these results would be either temperature dependent damping of  Alfv\'en waves or measurements along different coronal structures
formed at different temperatures. Possible temperature dependent damping may suggest some important role of thermal conduction in the damping of Alfv\'en waves
in the lower corona. This may even suggest some non-linear coupling between Alfv\'en and slow MHD modes \citep[see][]{2006A&A...456L..13Z}.
We believe this to be an important result as this will provide more insight in to the  dissipation mechanism of Alfv\'en waves.
Recent 3-D MHD models of \citet{2011ApJ...736....3V} explained role of Alfv\'en wave turbulence in the heating of solar chromosphere and corona.
They predicted velocity amplitude of Alfv\'en waves in the corona to be in the range of 20-40 km~s$^{-1}$ so as to maintain the typical active region loops.
In our analysis, we found almost similar wave velocity amplitude in the active region. Observed damping rate of Alfv\'en wave energy flux with 
 height is similar or slightly less than to the requirements of coronal active region. \citet{2014ApJ...786...28A} 
also measured non-thermal velocities in the range of 25-45 km~s$^{-1}$  using observation from EIS/Hinode along the on-disk
 individual coronal loop length. Their findings were consistent with the predictions from Alfv\'en wave turbulence model. However, 
 we would also like to point out that model of \citet{2011ApJ...736....3V} still do not include effects of thermal conduction and radiative losses.
 Therefore, at present, exact form of any relation between damping length of Alfv\'en wave turbulence and temperature can not be comprehended.
  Henceforth, these results demand for development of more sophisticated   3-D MHD models of Alfv\'en wave propagation and dissipation
  including the effects of thermal conduction and non-linear coupling between various MHD modes in the solar atmosphere.

\acknowledgments
Author thanks the referee for the careful reading and constructive criticism that helped improve the paper.
GRG is supported through the INSPIRE Faculty Award of Department of Science and Technology (DST), India. 
Author thanks T.~V. Zaqarashvili for the helpful discussion and P. Young for helpful clarifications.
Hinode is a Japanese mission developed and launched by ISAS/JAXA, collaborating with NAOJ as a domestic partner, 
NASA and STFC (UK) as international partners. Scientific operation of the Hinode mission is conducted by the Hinode science team
organized at ISAS/JAXA. This team mainly consists of scientists from institutes in the partner countries. Support for the post-launch 
operation is provided by JAXA and NAOJ (Japan), STFC (U.K.), NASA (U.S.A.), ESA, and NSC (Norway).

\end{document}